
\input amstex
\magnification=1200
\documentstyle{amsppt}
\NoRunningHeads
\NoBlackBoxes

\define\sLtwo{\operatorname{sl}(2,\Bbb C)}
\define\sUtwo{\operatorname{su}(2)}

\define\real{\operatorname{Re}}
\topmatter
\title
Topics in nonhamiltonian (magnetic--type) interaction of classical hamiltonian
dynamic systems
\endtitle
\author Denis Juriev\footnote{A preliminary electronic version of
this manuscript is located in the Texas e-print archive on mathematical physics
and has a number "mp\_arc/94-136" (May, 1994).\newline} \endauthor
\abstract\tenpoint A convenient algebraic structure to describe some forms
of dynamics of two hamiltonian systems with nonpotential (magnetic--type)
interaction is considered. An algebraic mechanism of generation of such
dynamics is explored on simple "toy" examples and models. Nonpotential chains
and their continuum limits are also considered. Examples of hybrid couplings
with both potential and nonpotential (magnetic--type) interactions are
discussed.
\endabstract
\endtopmatter
\document
\head I. Introduction\endhead

The evolution of hamiltonian systems (defined by Poisson brackets and
hamiltonians) attracts a lot of attention (see f.e. [1]). Many of such
systems are associated with Lie algebras, in this case Poisson brackets (the
so--called Lie--Berezin brackets) have a linear form and are restored from
the commutator in the Lie algebra [2] (it should be marked that nonlinear
Poisson brackets are also of interest [3]). The introducing of external fields
(which interact with the system nonpotentially, in general) may be accounted
by a transformation of commutator Lie algebras into the so-called isocommutator
Lie algebras (their $Q$--operator deformations). It means that if a Lie algebra
is realized by operators with an ordinary commutator $[X,Y]=XY-YX$, the
corresponding isocommutator Lie algebra (its $Q$--operator deformation) is
realised by operators with the so-called isocommutator $[X,Y]_Q=XQY-YQX$, where
an additional fixed operator $Q$ decodes an information on the external field
(see f.e. [4]). If the external field possesses a symmetry governed by the Lie
algebra $\frak g$, then it is rather convenient to describe the systems by
Lie $\frak g$--bunches [5]. This is a standard situation for magnetic-type
external fields (see f.e. [6]). By use of Lie $\frak g$--bunches one may
construct various hamiltonian systems nonpotentially controlled by external
fields, which dynamics maybe also hamiltonian, in particular. Such picture is
rather realistic if one supposes that a counteraction of the system on the
external fields maybe neglected. Nevertheless, it is not so in many important
cases. Therefore, it is interesting to consider a dynamics of two hamiltonian
systems, which interact with each other noncanonically (nonpotentially). For
example, a system of two charged spinning particles with nonpotential
magnetic--type interaction may be considered as an example. A convenient
algebraic structure to describe some forms of such dynamics seems to be of
the following definition (see, however, [7] for more general one).

\definition{Definition 1} The pair $(V_1,V_2)$ of linear spaces is called {\it
an isotopic pair\/} iff there are defined two mappings
$m_1:V_2\otimes\bigwedge^2V_1\mapsto V_1$ and
$m_2:V_1\otimes\bigwedge^2V_2\mapsto V_2$ such that the mappings $(X,Y)\mapsto
[X,Y]_A=m_1(A,X,Y)$ ($X,Y\in V_1$, $A\in V_2$) and $(A,B)\mapsto
[A,B]_X=m_2(X,A,B)$ ($A,B\in V_2$, $X\in V_1$) obey the Jacobi identity for
all values of a subscript parameter (such operations will be called {\it
isocommutators\/} and the subscript parameters will be called {\it
isotopic elements\/}) and are compatible to each other, i.e. the identities
$$\align
[X,Y]_{[A,B]_Z}=&\tfrac12([[X,Z]_A,Y]_B+[[X,Y]_A,Z]_B+[[Z,Y]_A,X]_B-\\
&[[X,Z]_B,Y]_A-[[X,Y]_B,Z]_A-[[Z,Y]_B,X]_A)\endalign$$
and
$$\align
[A,B]_{[X,Y]_C}=&\tfrac12([[A,C]_X,B]_Y+[[A,B]_X,C]_Y+[[C,B]_X,A]_Y-\\
&[[A,C]_Y,B]_X-[[A,B]_Y,C]_X-[[C,B]_Y,A]_X)\endalign$$
($X,Y,Z\in V_1$,
$A,B,C\in V_2$) hold.
\enddefinition

This definition is a result of an axiomatization of the following
construction. Let's consider an associative algebra $\Cal A$ (f.e. any matrix
algebra) and two linear subspaces $V_1$ and $V_2$ in it such that $V_1$ is
closed under the isocommutators $(X,Y)\mapsto [X,Y]_A=XAY-YAX$ with isotopic
elements $A$ from $V_2$, whereas $V_2$ is closed under the isocommutators
$(A,B)\mapsto [A,B]_X=AXB-BXA$ with isotopic elements $X$ from $V_1$. If a
family (linear space) $V_1$ of operators forms an isocommutator algebra with
a family (linear space) $V_2$ of admissible isotopic elements then $(V_1,V_2)$
is an isotopic pair via the so--called "isotopic duality" [5] (which maybe
regarded as a certain "algebraic manifestation" of the Third Law of Classical
Dynamics).

It should be mentioned that the isocommutators define families of Poisson
brackets $\{\cdot,\cdot\}_A$ and $\{\cdot,\cdot\}_X$ ($A\in V_2$, $X\in V_1$)
in the spaces $S^{\cdot}(V_1)$ and $S^{\cdot}(V_2)$, respectively.

\definition{Definition 2A} Let's consider two elements $\Cal H_1$ and $\Cal
H_2$ (hamiltonians) in $S^{\cdot}(V_1)$ and $S^{\cdot}(V_2)$, respectively.
The equations $$\dot X_t=\{\Cal H_1,X_t\}_{A_t},\quad \dot A_t=\{\Cal
H_2,A_t\}_{X_t},$$ where $X_t\in V_1$ and $A_t\in V_2$ will be called {\it
the\/} ({\it nonlinear\/}) {\it dynamical equations associated with the
isotopic pair $(V_1,V_2)$ and hamiltonians $\Cal H_1$ and $\Cal H_2$\/}.
\enddefinition

\definition{Definition 2B} Let's consider two elements $\Omega^{\pm}$ in
$V_1$ and $V_2$, respectively. The equations $$\dot X_t=[\Omega^+,X_t]_{A_t},
\quad \dot A_t=[\Omega^-,A_t]_{X_t},$$ where $X_t\in V_1$ and $A_t\in V_2$
will be called {\it the\/} ({\it nonlinear\/}) {\it dynamical equations\/}
({\it Euler formulas\/}) {\it associated with the isotopic pair $(V_1,V_2)$
and elements $\Omega^{\pm}$}.
\enddefinition

Let's consider several simple but crucial examples now. We shall treat the
subject purely mathematically and shall not specify its concrete physical
implications of a possible interest in details.

\head II. Noncanonically coupled rotators and Euler--Arnold tops \endhead

\subhead 2.1. Nonlinear integrable dynamics of noncanonically coupled rotators
\endsubhead Let's consider the Lie $\sLtwo$--bunch in $\pi_1$ (the adjoint
representation of $\sLtwo$)$^5$. It is defined by the next formulas

\

\centerline{(1)$\quad$
$\aligned
[m_{-1},m_0]_{l_0}&=m_{-1}\\
[m_1,m_{-1}]_{l_0}&=0\\
[m_1,m_0]_{l_0}&=m_1
\endaligned$
$\quad$
$\aligned
[m_{-1},m_0]_{l_{-1}}&=0\\
[m_1,m_{-1}]_{l_{-1}}&=2m_{-1}\\
[m_1,m_0]_{l_{-1}}&=2m_0
\endaligned$
$\quad$
$\aligned
[m_{-1},m_0]_{l_1}&=2m_0\\
[m_1,m_{-1}]_{l_1}&=-2m_1\\
[m_1,m_0]_{l_1}&=0
\endaligned$}

\ \newline
where $l_i$ ($i=-1,0,1$) are generators of $\sLtwo$
($[l_i,l_j]=(i-j)l_{i+j}$), $m_i$ ($i=-1,0,1$) form a basis in $\pi_1$
($l_i(m_j)=(i-j)m_{i+j}$).

As it was marked in [5] the operators $l_i$ form an isocommutator Lie algebra
with respect to isotopic elements from $\pi_1$ via the "isotopic duality",
namely

\

\centerline{(2)$\quad\qquad$
$\aligned
[l_{-1},l_0]_{m_0}&=-l_{-1}\\
[l_1,l_{-1}]_{m_0}&=0\\
[l_1,l_0]_{m_0}&=-l_1
\endaligned$
$\quad$
$\aligned
[l_{-1},l_0]_{m_{-1}}&=0\\
[l_1,l_{-1}]_{m_{-1}}&=-2l_{-1}\\
[l_1,l_0]_{m_{-1}}&=-2l_0
\endaligned$
$\quad$
$\aligned
[l_{-1},l_0]_{m_1}&=-2l_0\\
[l_1,l_{-1}]_{m_1}&=2l_1\\
[l_1,l_0]_{m_1}&=0
\endaligned$}

\

It can be easily verified that isocommutators (1) and (2) define a structure
of an isotopic pair in $\pi_1\oplus\pi_1$. Mark that all isocommutators are
$r$--matrix ones, i.e. may be constructed from a standard Lie bracket in
$3$--dimensional Lie algebra $\operatorname{\frak s\frak l}(2,\Bbb R)$ by use
of classical $r$--matrices (see [8]).

Let's denote the first summand $\pi_1$ by $\pi_1^+$ and the second summand by
$\pi_1^-$. Let's also consider two fixed elements $\Omega^{\pm}$ in
$\pi_1^{\pm}$, respectively
($\Omega^+=\Omega^+_{-1}m_{-1}+\Omega^+_0m_0+\Omega^+_1m_1$,
$\Omega^-=\Omega^-_{-1}l_{-1}+\Omega^-_0l_0+\Omega^-_1l_1$), and two variables
$A$ and $B$ from $\pi_1^+$ and $\pi_1^-$ ($A=A_{-1}m_{-1}+A_0m_0+A_1m_1$ and
$B=B_{-1}l_{-1}+B_0l_0+B_1l_1$). The dynamical equations (Euler formulas)
defined by $\Omega^{\pm}$ will have the form
$$\left\{\aligned
\dot A_{-1}&=
\Omega^+_{-1}(A_0B_0-2A_1B_{-1})-\Omega^+_0A_{-1}B_0+2\Omega^+_1A_{-1}B_{-1}\\
\dot A_0&=
2\Omega^+_{-1}A_0B_1+2\Omega^+_0(A_1B_{-1}-A_{-1}B_1)+2\Omega^+_1A_0B_{-1}\\
\dot A_1&=
2\Omega^+_{-1}A_1B_1-\Omega^+_0A_1B_0+\Omega^+_1(A_0B_0-2A_{-1}B_1)
\endaligned\right.$$
$$\left\{\aligned
\dot B_{-1}&=
-\Omega^-_{-1}(B_0A_0-2B_1A_{-1})+\Omega^-_0B_{-1}A_0-2\Omega^-_1B_{-1}A_{-1}\\
\dot B_0&=
-2\Omega^-_{-1}B_0A_1-2\Omega^-_0(B_1A_{-1}-B_{-1}A_1)-2\Omega^+_1B_0A_{-1}\\
\dot B_1&=
-2\Omega^-_{-1}B_1A_1+\Omega^-_0B_1A_0-\Omega^-_1(B_0A_0-2B_{-1}A_1)
\endaligned\right.
$$

It is rather convenient to consider the compact real form of the isotopic
pair $(\pi_1^+,\pi_1^-)$. Let's denote
$$\align l_x=\tfrac{i}2(l_1-l_{-1}),\quad l_y=&\tfrac12(l_1+l_{-1}),\quad
l_z=i l_0\\
m_x=\tfrac{i}2(m_1-m_{-1}),\quad m_y=&\tfrac12(m_1+m_{-1}),\quad
m_z=i m_0.\endalign$$

The dynamical equation maybe rewritten as
$$\left\{\aligned
\dot A_x&=-\Omega^+_x(A_yB_y+A_zB_z)+\Omega^+_yA_xB_y+\Omega^+_zA_xB_z\\
\dot A_y&=-\Omega^+_y(A_xB_x+A_zB_z)+\Omega^+_xA_yB_x+\Omega^+_zA_yB_z\\
\dot A_z&=-\Omega^+_z(A_xB_x+A_yB_y)+\Omega^+_xA_zB_x+\Omega^+_yA_zB_y
\endaligned\right.$$
$$\left\{\aligned
\dot B_x&=-\Omega^-_x(B_yA_y+B_zA_z)+\Omega^-_yB_xA_y+\Omega^-_zB_xA_z\\
\dot B_y&=-\Omega^-_y(B_xA_x+B_zA_z)+\Omega^-_xB_yA_x+\Omega^-_zB_yA_z\\
\dot B_z&=-\Omega^-_z(B_xA_x+B_yA_y)+\Omega^-_xB_zA_x+\Omega^-_yB_zA_y
\endaligned\right.$$
after the change $\Omega^-\to-\Omega^-$. They maybe also written in a more
compact form
$$\left\{\aligned
\dot A&=-\left<A,B\right>\Omega^++\left<\Omega^+,B\right>A\\
\dot B&=-\left<A,B\right>\Omega^-+\left<\Omega^-,A\right>B
\endaligned\right. $$
here brackets $\left<\cdot,\cdot\right>$ denote $\sUtwo$--invariant inner
products in $\pi_1^{\pm}$. Such nonlinear dynamics maybe interpreted as one
for the noncanonically coupled rotators. It possesses two trivial
integrals of motion $K=\left<A,B\right>$ and
$L=\left<A,\Omega^-\right>-\left<B,\Omega^+\right>$.

Let's denote
$\left<\Omega^+,B\right>+\frac12L=\left<\Omega^-,A\right>-\frac12L$ by $y$.
Then
$$\left\{\aligned
\dot A&=-K\Omega^++(y-\tfrac12L)A\\
\dot B&=-K\Omega^-+(y+\tfrac12L)B
\endaligned\right.
$$
whereas
$$\dot y=y^2-(\tfrac14L^2+K\beta),$$
$\beta=\left<\Omega^+,\Omega^-\right>$, so the dynamical equations (Euler
formulas) are integrable in elementary functions.

\subhead 2.2. Nonlinear integrable dynamics of noncanonically coupled
Euler--Arnold tops \endsubhead Noncanonically coupled Euler--Arnold tops are
realized by means of the same isotopic pair as noncanonically coupled rotators
but with quadratic hamiltonians $\Cal H_1=\left<A,T_1A\right>$ and $\Cal
H_2=\left<B,T_2B\right>$ ($T_i=T^*_i$). The corresponding dynamical equations
maybe
written as
$$\left\{\aligned
\dot A&=-\left<T_1A,A\right>B+\left<T_1A,B\right>A\\
\dot B&=-\left<T_2B,A\right>B+\left<T_2B,B\right>A
\endaligned\right.
$$
Such system admits integrals of motion $K_A=\left<TA,A\right>$,
$K_B=\left<TB,B\right>$. If $T_1=T_2=T$ then
$L=\left<TA,B\right>=\left<TB,A\right>$ is also an integral; the equations
maybe rewritten as
$$\left\{\aligned
\dot A&=LA-K_AB\\
\dot B&=K_BA-LB
\endaligned\right.
$$
and are easily integrated.

It is rather interesting to obtain solutions of dynamical equations with
$T_1\ne T_2$ by use of some sort of "dynamical dressing transformations"
(cf. [9]).

\head III. Noncanonically coupled oscillators and their chains \endhead

\subhead 3.1. Nonlinear integrable dynamics of noncanonically coupled
oscillators \endsubhead Let's consider a standard representation of the
Heisenberg algebra with three generators $p$, $q$ and $r$ ($[p,q]=r$,
$[p,r]=[q,r]=0$) by $3\times3$ matrices:
$$p=\left(\matrix 0 & 1 & 0 \\ 0 & 0 & 0 \\ 0 & 0 & 0 \endmatrix\right),\quad
q=\left(\matrix 0 & 0 & 0 \\ 0 & 0 & 1 \\ 0 & 0 & 0 \endmatrix\right), \quad
r=\left(\matrix 0 & 0 & 1 \\ 0 & 0 & 0 \\ 0 & 0 & 0 \endmatrix\right).$$

The linear space of admissible isotopic elements is generated by three elements
$a$, $b$ and $c$, which are represented by matrices
$$a=\left(\matrix 0 & 0 & 0 \\ 1 & 0 & 0 \\ 0 & 0 & 0 \endmatrix\right), \quad
b=\left(\matrix 0 & 0 & 0 \\ 0 & 0 & 0 \\ 0 & 1 & 0 \endmatrix\right), \quad
c=\left(\matrix 0 & 0 & 0 \\ 0 & 1 & 0 \\ 0 & 0 & 0 \endmatrix\right).$$

The isocommutators have the form

\

\centerline{
$\aligned
[p,q]_a&=0\\
[p,r]_a&=r\\
[q,r]_a&=0
\endaligned$
$\quad$
$\aligned
[p,q]_b&=0\\
[p,r]_b&=0\\
[q,r]_b&=-r
\endaligned$
$\quad$
$\aligned
[p,q]_c=r\\
[p,r]_c=0\\
[q,r]_c=0
\endaligned$}

\

The elements $a$, $b$ and $c$ are closed themselves under isocommutators with
isotopic elements $p$, $q$ and $r$ via the "isotopic duality":

\

\centerline{
$\aligned
[a,b]_p&=0\\
[a,c]_p&=c\\
[b,c]_p&=0
\endaligned$
$\quad$
$\aligned
[a,b]_q&=0\\
[a,c]_q&=0\\
[b,c]_q&=-c
\endaligned$
$\quad$
$\aligned
[a,b]_r=c\\
[b,c]_r=0\\
[a,c]_r=0
\endaligned$}

\

The linear spaces generated by $p$, $q$, $r$ and $a$, $b$, $c$ form an
isotopic pair. One may associate noncanonically coupled oscillators with it;
namely, let's consider $p$, $q$, $r$ and $a$, $b$, $c$ as linear functionals
(we shall denote them by capitals) on dual spaces; hamiltonians $\Cal H_1$
and $\Cal H_2$ will be of the form $\Cal H_1=P^2+Q^2$ and $\Cal H_2=A^2+B^2$,
the dynamical equations will be written as

\

\centerline{
$\left\{\aligned
\dot Q&=2RPC\\
\dot P&=-2RQC\\
\dot R&=2PRA-2QRB
\endaligned\right.$
$\quad$
$\left\{\aligned
\dot A&=-2CBR\\
\dot B&=2CAR\\
\dot C&=2ACP-2BCQ
\endaligned\right.$}

\

It should be marked that hamiltonians are integrals of motion here, also
three integrals maybe written: $\Cal M=AQ-BP$, $\Cal N=BQ+AP$ and $\Cal
L=CR-BP-AQ=CR-2BP-\Cal M=CR-2AQ+\Cal M$. The presence of
five integrals essentially simplifies a picture, so an integration of the
system of two noncanonically coupled oscillators becomes an easy but
interesting exercise. Let's perform it. Put $\Cal H_1=h^2_1$, $\Cal
H^2_2=h^2_2$, $P=h_1\cos\varphi$, $Q=h_1\sin\varphi$, $A=h_2\cos\psi$,
$B=h_2\sin\psi$, then the condition $\dot{\Cal M}=\dot{\Cal N}=0$ gives
$\cos(\varphi-\psi)=-\frac{\Cal N}{h_1h_2}$, $\sin(\varphi-\psi)=\frac{\Cal
M}{h_1h_2}$, so $\vartheta:=\psi-\varphi=\arctan(\frac{\Cal M}{\Cal N})$. Also
$\dot\varphi=\dot\psi=2CR$, $\dot R=2Rh_1h_2\cos(\varphi+\psi)$, $\dot
C=2Ch_1h_2\cos(\varphi+\psi)$, hence $C=\varkappa R$ and
$\dot\varphi=2\varkappa R^2$, $\dot R=2h_1h_2R\cos(2\varphi+\vartheta)$. Let's
consider $R$ as a function of $\varphi$ then
$RR'_{\varphi}=h_1h_2\cos(2\varphi+\vartheta)$ and
$R=\sqrt{\frac{\Cal L}{\varkappa}+
\frac{h_1h_2}{\varkappa}\sin(2\varphi+\vartheta)}$ (to receive
this fact one may also use a conservation law $\dot{\Cal L}=0$). It should be
marked that the plane curve, which is defined by the equation $R=R(\varphi$)
in polar coordinates, is the Booth lemniscate. Substituting
the resulted expression for $R=R(\varphi)$ into the formula for $\dot\varphi$
one obtains that $\dot\varphi=
2\Cal L+2h_1h_2\sin(2\varphi+\vartheta)$.

\subhead 3.2. Dynamics of periodic and nonperiodic chains of noncanonically
coupled oscillators \endsubhead The dynamical equations for two noncanonically
coupled oscillators are immediately generalized on (periodic or infinite
nonperiodic) chains of them, namely
$$\left\{\aligned
\dot Q_i&=R_iP_i(R_{i-1}+R_{i+1})\\
\dot P_i&=-R_iQ_i(R_{i-1}+R_{i+1})\\
\dot R_i&=R_iP_i(P_{i-1}+P_{i+1})-R_iQ_i(Q_{i-1}+Q_{i+1})
\endaligned\right.$$

The hamiltonians $\Cal H_i=P^2_i+Q^2_i$ are certainly integrals of motion.
Put $\Cal H_i=h^2_i$, $P_i=h_i\cos\varphi_i$, $Q_i=h_i\sin\varphi_i$, then
$\dot\varphi_i=R_i(R_{i-1}+R_{i+1})$, $\dot R_i=R_ih_i(h_{i-1}\cos(\varphi_i+
\varphi_{i-1})+h_{i+1}\cos(\varphi_i+\varphi_{i+1}))$.

Let's introduce new "coupled" variables $\psi_i=\varphi_i+\varphi_{i+1}$ and
$S_i=R_iR_{i+1}$ as well as $H_i=h_ih_{i+1}$. The dynamical equations
are rewritten as follows
$$\left\{\aligned\dot\psi_i&=S_{i-1}+2S_i+S_{i+1}\\
\dot S_i&=S_i(H_{i-1}\cos\psi_{i-1}+2H_i\cos\psi_i+
H_{i+1}\cos\psi_{i+1})\endaligned\right.$$
Let's put $T_i=H_ie^{\sqrt{-1}\psi_i}$ now, then
$$\left\{\aligned\dot S_i&=\real (T_{i-1}+2T_i+T_{i+1})S_i\\
\dot T_i&=\sqrt{-1}(S_{i-1}+2S_i+S_{i+1})T_i\endaligned\right.$$

\subhead 3.3. A continuum limit of the noncanonically coupled oscillator chain
\endsubhead One may consider noncanonically coupled oscillator chain with
changed signs, i.e.
$$\left\{\aligned
\dot Q_i&=R_iP_i(R_{i+1}-R_{i-1})\\
\dot P_i&=-R_iQ_i(R_{i+1}-R_{i-1})\\
\dot R_i&=R_iP_i(P_{i+1}-P_{i-1})-R_iQ_i(Q_{i+1}-Q_{i-1})\endaligned\right.$$

Such chain admits a natural continuum (field) limit
$$\left\{\aligned
\dot q&=prr'_x\\
\dot p&=-qrr'_x\\
\dot r&=rpp'_x-rqq'_x
\endaligned\right.$$
$q\!=\!q(t,x)$, $p\!=\!p(t,x)$, $r\!=\!r(t,x)$. The function
$h^2\!=\!p^2\!+\!q^2$ is an integral of motion, so it is convenient to put
$p=h\cos\varphi$, $q\!=\!h\sin\varphi$. Then
$$\left\{\aligned\dot\phi&=rr'_x\\
\dot r&=-rh^2\sin{2\varphi}\varphi'_x
\endaligned\right.$$
Let's put $s\!=\!r^2$, $t\!=\!h^2e^{2\sssize\sqrt{-1}\tsize\varphi}$, then
$$\left\{\aligned\dot s&=-\ssize\real\dsize(t'_x)s\\
\dot t&=\ssize\sqrt{-1}\dsize s'_xt
\endaligned\right.$$

\head IV. Other examples \endhead

\subhead 4.1. Dynamic system connected with the isotopic pair of $3\times3$
symmetric and skew--symmetric matrices \endsubhead One of the most interesting
examples of isotopic pairs is related to $n\times n$ symmetric and
skew--symmetric matrices. Let's consider the simplest case $n=3$. It is rather
convenient to introduce the following basises:
$$l_z=\left(\matrix 0 & 1 & 0 \\ -1 & 0 & 0 \\ 0 & 0 & 0 \endmatrix\right),
\quad l_x=\left(\matrix 0 & 0 & 0 \\ 0 & 0 & 1 \\ 0 & -1 & 0
\endmatrix\right), \quad l_y=\left(\matrix 0 & 0 & 1 \\ 0 & 0 & 0 \\ -1 & 0 &
0 \endmatrix\right)$$
and
$$\align
&m_{xy}=\left(\matrix 0 & 1 & 0 \\ 1 & 0 & 0 \\ 0 & 0 & 0 \endmatrix\right),
\quad m_{yz}=\left(\matrix 0 & 0 & 0 \\ 0 & 0 & 1 \\ 0 & 1 & 0
\endmatrix\right), \quad m_{xz}=\left(\matrix 0 & 0 & 1 \\ 0 & 0 & 0 \\ 1 & 0
& 0 \endmatrix\right),\\
&\\
&m_{xx}=\left(\matrix 2 & 0 & 0 \\ 0 & 0 & 0 \\ 0 & 0 & 0 \endmatrix\right),
\quad m_{yy}=\left(\matrix 0 & 0 & 0 \\ 0 & 2 & 0 \\ 0 & 0 & 0
\endmatrix\right), \quad m_{zz}=\left(\matrix 0 & 0 & 0 \\ 0 & 0 & 0 \\ 0 & 0
& 2 \endmatrix\right).\endalign$$
Elements $l_a$ form a basis in the space of skew--symmetric matrices, whereas
$w_{ab}$ form a basis in the space of symmetric matrices. The isocommutators
have the form
$$\align
[m_{ab},m_{cd}]_{l_e}&=\epsilon_{ace}m_{bd}+\epsilon_{ade}m_{bc}+
\epsilon_{bce}m_{ad}+\epsilon_{bde}m_{ac},\\
[l_a,l_b]_{m_{cd}}&=\epsilon_{abc}l_d+\epsilon_{abd}l_c,\endalign$$
here $\epsilon_{abc}$ is a totally antisymmetric tensor.

Let's fix a skew--symmetric matrix $\Omega^-$ and a symmetric matrix
$\Omega^+$. If the linear functionals defined by $l_a$ and $m_{ab}$ in the
dual spaces are denoted by capitals $L_a$ and $M_{ab}$, respectively, then
the dynamical equations (Euler formulas) will have the form
$$\left\{\aligned \dot L_a&=\epsilon_{bcd}\Omega^-_bL_cM_{ad}\\
\dot M_{ab}&=\epsilon_{cde}\Omega^+_{ac}M_{bd}L_e+
\epsilon_{cde}\Omega^+_{bc}M_{ad}L_e\endaligned\right.$$

Unfortunately, I do not know whether this system is integrable.

\subhead 4.2. The hybrid coupling: "elastoplastic" spring \endsubhead It is
rather interesting to consider the case then two hamiltoinian systems are
coupled by interaction terms in a hamiltonian and simultaneously
noncanonically. Below we shall consider an "elastoplastic" string, which is a
hybrid of the ordinary "elastic" spring with noncanonically coupled
oscillators.

The isotopic pair is the same as for noncanonically coupled oscillators but
the hamiltonian is of the form $\Cal H=\Cal H_1+\Cal H_2-2QB-2PA$; the
dynamical equations have the form

\

\centerline{
$\left\{\aligned
\dot Q&=2RC(P-A)\\
\dot P&=-2RC(Q-B)\\
\dot R&=-2(P-A)^2R+2(Q-B)^2R
\endaligned\right.$
$\quad$
$\left\{\aligned
\dot A&=2RC(Q-B)\\
\dot B&=-2RC(P-A)\\
\dot C&=-2(P-A)^2C+2(Q-B)^2C
\endaligned\right.$}

\

Let's denote $D=P-A$, $G=Q-B$, then $\Cal J^2=D^2+G^2$ is an integral of
motion so it is convenient to put $D=\Cal J\sin\varphi$, $G=\Cal
J\cos\varphi$. Moreover, $C=\lambda R$ and $\dot R=2\Cal
J^2R^2\cos{2\varphi}$, whereas $\dot\varphi=-4\lambda R^2$. Hence,
$-2\lambda RR'_\varphi=\Cal J^2\cos{2\varphi}$ and $R=\sqrt{\frac{\Cal
L}\lambda-\frac1{2\lambda}\Cal J^2\sin{2\varphi}}$ (and the corresponding
plane curve is the Booth lemniscate again), whereas $\dot\varphi=-4\Cal
L+2\Cal J^2\sin{2\varphi}$.

\subhead 4.3. "Elastoplastic" spring with general interaction potential
\endsubhead The described picture is straightforwardly generalized on
arbitrary interaction potentials. Namely, let's consider the hamiltonian of
the form $\Cal H=(P-A)^2+V(Q-B)$, where $V$ is an interaction potential. The
dynamical equations are written in the form
$$\left\{\aligned
\dot Q&=2RC(P-A)\\
\dot P&=-RCV'(Q-B)\\
\dot R&=-2(P-A)^2R+(Q-B)V'(Q-B)R
\endaligned\right.$$
$$\left\{\aligned
\dot A&=RCV'(Q-B)\\
\dot B&=-2RC(P-A)\\
\dot C&=-2(P-A)^2C+(Q-B)V'(Q-B)C
\endaligned\right.$$

Let's denote $D\!=\!P\!-\!A$, $G\!=\!Q\!-\!B$; $\Cal J^2\!=\!D^2\!+\!V(G)$ is
an integral of motion so it is convenient to put $D\!=\!\Cal J\sin\varphi$,
$G\!=\!V^{-1}(\Cal J^2\cos^2\varphi)$. Moreover $C\!=\!\lambda R$ and
$\dot R\!=\!\left[-2\Cal J^2\sin^2\varphi\!+
\!V^{-1}(\Cal J^2\cos^2\varphi)V'(V^{-1}(\Cal J^2\cos^2\varphi))\right]R$,
whereas $\dot\varphi\!=\!-\frac{2\lambda R^2}{\Cal
J\cos\varphi}\times V'(V^{-1}(\Cal J^2\cos^2\varphi))$. Hence $-2\lambda
RR'_\varphi
V'(V^{-1}(\Cal J^2\cos^2\varphi))\!=\!\Cal J\cos\varphi\left[-2\Cal
J^2\sin^2\varphi\right.+\left.V^{-1}(\Cal J^2\cos^2\varphi)V'(V^{-1}(\Cal
J^2\cos^2\varphi))\right]$ and
$$R=\sqrt{\frac1{\lambda}\int\Cal J\cos\varphi
\frac{2\Cal J^2\sin^2\varphi-V^{-1}(\Cal J^2\cos^2\varphi)V'(V^{-1}(\Cal
J^2\cos^2\varphi))}{V'(V^{-1}(\Cal J^2\cos^2\varphi))}d\varphi},$$ whereas
$$\dot\varphi=\frac2{\cos\varphi}\int\cos\varphi\left[V^{-1}(\Cal
J^2\cos^2\varphi)-\frac{2\Cal J^2\sin^2\varphi}{V'(V^{-1}(\Cal
J^2\cos^2\varphi))}\right]d\varphi.$$

\head V. Conclusions \endhead

Thus, a certain convenient algebraic structure (a structure of {\it isotopic
pair\/}) to describe some forms of classical dynamics of two hamiltonian
systems
with nonpotential (magnetic--type) interaction was considered. An algebraic
mechanism of generation of such dynamics was explored on simple "toy" examples
(coupled rotators, tops and oscillators). It is analogous to well-known one for
hamiltonian systems constructed from Lie algebras (see f.e. [10]). The
nonpotential chains and their continuum (field) limits were also considered.
Examples of hybrid couplings with both potential and nonpotential terms were
discussed.

\Refs
\tenpoint
\roster
\item"[1]" Arnold V.I., {\it Mathematical methods of classical mechanics},
Springer--Verlag, 1980; Dubro\-vin B.A., Novikov S.P., Fomenko A.T., {\it
Modern
geometry -- methods and applications}, Springer--Verlag, 1988; Fomenko A.T.,
{\it Symplectic geometry. Methods and applications}, Gordon \& Beach, 1988;
Arnold V.I., Kozlov V.V., Neustadt A.I., Current Problems of Math., Fundamental
Directions. Vol.3. Moscow, VINITI, 1985, pp.5-303 [in Russian].
\item"[2]" Kirillov A.A., {\it Elements of the theory of representations},
Springer--Verlag, 1976; Perelomov A.M., {\it Integrable systems of classical
mechanics and Lie algebras}, Birkhauser--Verlag, 1990.
\item"[3]" Karasev M.V., Maslov V.P., {\it Nonlinear Poisson brackets:
geometry and quantization}, Amer. Math. Soc., Providence, RI, 1993.
\item"[4]" Santilli R.M., {\it Foundations of theoretical mechanics. II.
Birkhoffian generalization of Ha\-miltonian mechanics}, Springer--Verlag, 1982.
\item"[5]" Juriev D., Topics in hidden symmetries, E--print (LANL Electronic
Archive on Theor. High Energy Phys.): {\it hep-th/9405050} (1994).
\item"[6]" Dubrovin B.A., Krichever I.M., Novikov S.P., Current Problems of
Math., Fundamental Directions. Vol.4. Moscow, VINITI, 1985, pp.179-284 [in
Russian]; Thirring W., {\it Classical dynamical systems}, Springer--Verlag,
1978.
\item"[7]" Juriev D., On the nonhamiltonian interaction of two rotators,
E--print (MSRI Electronic Archive on Diff. Geom. and Global Anal.): {\it
dg-ga/9409005\/} (1994).
\item"[8]" Semenov-Tian-Shansky M.A., Funct. Anal. Appl. 17, 259 (1983),
Zap. Na\-uchn. Sem. LOMI 133, 228 (1984).
\item"[9]" Semenov-Tian-Shansky M.A., Publ. RIMS 21, 1237 (1985), Zap. Nauchn.
Sem. LOMI 150, 119 (1986).
\item"[10]" Fomenko A.T., Trofimov V.V., {\it Geometric and algebraic
mechanisms
of integrability of hamiltonian systems on homogeneous spaces and Lie
algebras}.
In "Current Probl. Math.: Fundamental Directions 16", 1987, pp.227-295.
\endroster
\endRefs
\enddocument